\documentclass{elsart3}
\usepackage{graphicx}
\usepackage{amssymb}
\begin{document}

\begin{frontmatter}
\title{Susceptibility inhomogeneity and non-Fermi liquid behavior\\ in UCu$_{5-x}$Pt$_x$}
\author[UCR]{D.~E. MacLaughlin\corauthref{MacL}}
\corauth[MacL]{Tel. +1-951-827-5344, Fax +1-951-827-4529, e-mail: macl@physics.ucr.edu.}
\author[UCR]{M.~S. Rose}
\author[CSULA]{O.~O. Bernal}
\author[LANL,JAERI]{R.~H. Heffner}
\author[KOL]{\ G.~J. Nieuwenhuys}
\author[UCSD]{R. Chau}
\author[UCSD]{M.~B. Maple}
\address[UCR]{Department of Physics, University of California, Riverside, CA 92521, U.S.A.}
\address[CSULA]{Department of Physics and Astronomy, California State University, Los Angeles, CA 90032, U.S.A.}
\address[LANL]{Los Alamos National Laboratory, K764, Los Alamos, NM 87545, U.S.A.}
\address[JAERI]{Japan Atomic Energy Research Institute, Tokai, Ibaraki-ken, 319-1195, Japan}
\address[KOL]{Kamerlingh Onnes Laboratory, Leiden University, 2300 RA Leiden, The Netherlands}
\address[UCSD]{Department of Physics and Institute for Pure and Applied Physical Sciences, University of California, San Diego, La Jolla, California 92093, U.S.A.}

\begin{abstract}
Transverse-field $\mu$SR shifts and relaxation rates have been measured in the non-Fermi liquid (NFL) alloy system~UCu$_{5-x}$Pt$_x$, $x = 1.0$, 1.5, and 2.5. At low temperatures the fractional spread in Knight shifts~$\delta K/K \approx \delta\chi/\chi$ is $\gtrsim 2$ for $x = 1$, but is only half this value for $x = 1.5$ and 2.5. 
In a disorder-driven scenario where the NFL behavior is due to a broadly distributed (Kondo or Griffiths-phase cluster) characteristic energy~$E$, our results indicate that $\delta E/E_{\rm av} \approx (\delta K/K)_{T\to0}$ is similar for UCu$_{5-x}$Pd$_x$ ($x = 1$ and 1.5) and UCu$_4$Pt, but is reduced for UCu$_{5-x}$Pt$_x$, $x = 1.5$ and 2.5. This reduction is due to a marked increase of $E$ with increasing $x$; the spread~$\delta E$ is found to be roughly independent of $x$. Our results correlate with the observed suppression of other NFL anomalies for $x > 1$ in UCu$_{5-x}$Pt$_x$ but not in UCu$_{5-x}$Pd$_x$, and are further evidence for the importance of disorder in the NFL behavior of both these alloy systems. 
\end{abstract}

\end{frontmatter}

\section{Introduction}
UCu$_5$ is a heavy-fermion antiferromagnet (N\'eel temperature~$T_N \approx 16$~K)~\cite{ORFF85}. In the alloy systems~UCu$_{5-x}$Pd$_x$ and UCu$_{5-x}$Pt$_x$ $T_N$ is suppressed to 0 in the neighborhood of $x = 1$, beyond which point non-Fermi liquid (NFL) behavior is observed in both systems. For UCu$_{5-x}$Pt$_x$ NFL signatures are strong only for $x({\rm Pt}) \approx 1$~\cite{ChMa96,CFM95}, whereas for UCu$_{5-x}$Pd$_x$ NFL behavior is significant over a range of concentrations ($1 \lesssim x({\rm Pd}) \lesssim 2$). This raises the possibility that NFL behavior is strong only in UCu$_4$Pt because in this alloy series, unlike UCu$_{5-x}$Pd$_x$, the concentration~$x({\rm Pt}) \approx 1$ is a quantum critical point~\cite{VNvS02}.

In these alloys the transverse-field $\mu$SR (TF-$\mu$SR) relaxation rate (i.e., the width of the $\mu$SR line) is due to an inhomogeneous spread $\delta K$ of muon Knight shifts~$K$~\cite{BMAF96,MHBI04}.  This spread is in turn a measure of the spread~$\delta\chi$ in magnetic susceptibility~$\chi$ from the basic relation~$K \propto \chi$. The relaxation rate becomes very large at low temperatures in a number of NFL materials including UCu$_{5-x}$Pd$_x$ and UCu$_4$Pt~\cite{MHBI04}. Early NMR observations of this effect in UCu$_{5-x}$Pd$_x$~\cite{BMLA95} were taken as evidence for  the ``Kondo disorder'' model~\cite{MDK96,MDK97} for disorder-driven NFL behavior. This motivated a similar TF-$\mu$SR study of the UCu$_{5-x}$Pt$_x$ alloy system.

\section{Experimental Results}
The $\mu$SR experiments reported here were carried out using the General Purpose Spectrometer at the Paul Scherrer Institute, Villigen, Switzerland. 
Measured TF-$\mu$SR rates in UCu$_{5-x}$Pt$_x$ , $x = 1.0$, 1.5, and 2.5, and UCu$_4$Pd are shown in Fig.~1. 
 \begin{figure}[ht]
     \begin{center}
     \includegraphics[width=0.45\textwidth]{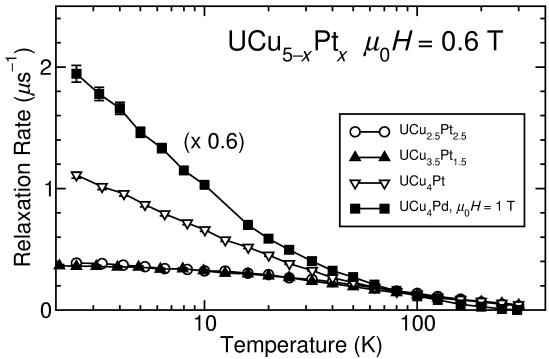}
     \end{center}
     \caption{Temperature dependence of the transverse-field exponential muon spin relaxation rate in UCu$_{5-x}$Pt$_x$ , $x = 1.0$, 1.5, and 2.5, and UCu$_4$Pd.}
 \end{figure}
The increase of exponential relaxation rate with decreasing temperature in UCu$_4$Pt is found to be about 40\% less than that in UCu$_4$Pd after correction for the difference in measuring fields (the relaxation rate, which is due to the inhomogeneous magnetization, is proportional to field). The increase is weaker still for UCu$_{5-x}$Pt$_x$, $x = 1.5$ and 2.5, where the rates are practically independent of $x$. 

A heuristic model~\cite{MHBI04} is useful in understanding this low-temperature increase. Assume a Curie-Weiss form 
\begin{equation} \chi(T) = C/(T + E) \,, \label{eq:curie} \end{equation}
where $E$ is a characteristic energy (here $k_B = 1$) that is inhomogeneously distributed. Then crudely (to first order) a spread $\delta E$ in $E$ yields a spread 
\begin{equation} \delta\chi = (\delta E/C) \chi_{\rm av}^2 \end{equation}
in $\chi$, where $\chi_{\rm av}$ is the spatially-averaged susceptibility. So we expect 
\begin{equation} \delta\chi/\chi_{\rm av} = (\delta E/C) \chi_{\rm av}
\label{eq:bernal}
\end{equation}
if disorder in the characteristic energy produces an inhomogeneous distribution of the susceptibility. 

A plot of the fractional spread in muon Knight shift $\delta K/(a^{*}\chi_{\rm av})$ vs.\ $\chi_{\rm av}$, with temperature an implicit parameter, is a useful presentation of these results. Here $a^{*}$ is an effective coupling constant that describes the interaction between $f$-ion spins and muons taking into account the spatial correlation of the disorder~\cite{BMAF96,MBL96}. Then $\delta K/(a^{*}\chi_{\rm av})$ is essentially the fractional spread $\delta\chi/\chi_{\rm av}$ in the inhomogeneous susceptibility. According to Eq.~(\ref{eq:bernal}) $\delta K/(a^{*}\chi_{\rm av})$ should be roughly proportional to $\chi_{\rm av}$, with slope~$\delta E/C$. 

The results for UCu$_{5-x}$Pt$_x$, $x = 1.0$, 1.5, and 2.5, and UCu$_4$Pd, shown in Fig.~2, exhibit this behavior (the small feature for $x = 2.5$ is mainly due to a low-temperature anomaly in $\chi$, probably caused by paramagnetic impurities). 
 \begin{figure}[ht]
     \begin{center}
     \includegraphics[width=0.45\textwidth]{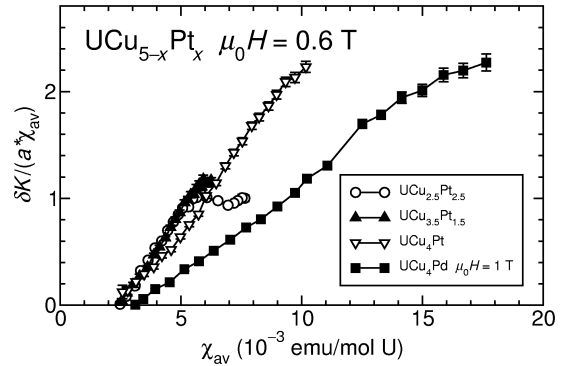}
     \end{center}
     \caption{Plots of fractional Knight shift spread~$\delta K/(a^{*}\chi_{\rm av})$ vs.\ spatially-averaged (bulk) susceptibility $\chi_{\rm av}$, with temperature an implicit parameter, in UCu$_{5-x}$Pt$_x$ , $x = 1.0$, 1.5, and 2.5, and UCu$_4$Pd.}
 \end{figure}
Here $a^{*} = 0.096\ \rm mol\ emu^{-1}$ has been calculated assuming dipolar coupling, as is the case in UCu$_{5-x}$Pd$_x$ to a good approximation~\cite{BMAF96}. Experimentally $\delta K/(a^{*}\chi_{\rm av})$ is not quite proportional to $\chi$, which is natural given the crude linear approximation of Eq.~(\ref{eq:bernal}). The departure from proportionality at small $\chi$ (high temperatures) may be due to thermally-activated muon diffusion, which motionally narrows the line and reduces $\delta K$, but this mechanism is activated and cannot account for the gradual and roughly proportional increase of $\delta K/(a^{*}\chi_{\rm av})$ with increasing $\chi_{\rm av}$ at lower temperatures. This behavior, which as outlined above provides evidence for a broad distribution of characteristic energies, is observed for all samples.

\section{Discussion}
We observe large values of $\delta K/(a^{*}\chi_{\rm av})$ ($\gtrsim$2) at low temperatures in both UCu$_4$Pt and UCu$_4$Pd. Since $\chi_{\rm av}(0) \approx C/E_{\rm av}$, where $E_{\rm av}$ is the spatially-averaged characteristic energy, we have 
\begin{equation}
\delta K/(a^{*}\chi_{\rm av})_{T=0} \approx (\delta\chi/\chi_{\rm av})_{T=0} = \delta E/E_{\rm av}
\label{eq:lowtemp}
\end{equation} 
from Eqs.~(\ref{eq:curie})--(\ref{eq:bernal}). Thus we find a broad distribution of $E$ ($\delta E \gtrsim 2 E_{\rm av}$) in the ``inhomogeneous characteristic energy'' scenario described above. Such a broad distribution is essential for NFL behavior in the disorder-driven Kondo disorder~\cite{BMLA95,MDK96,MDK97} and Griffiths-Phase~\cite{CNJ00} models, since in these pictures NFL properties are due to uncompensated spins with $E < T$~\cite{MHBI04,BMLA95,MBL96,nofit}.

In UCu$_{5-x}$Pt$_x$, $x = 1.5$ and 2.5, the values of $\delta K/(a^{*}\chi_{\rm av})$ as $T \to 0$ (i.e., the data for highest susceptibilites in Fig.~2) are smaller than for $x = 1.0$. This result agrees with thermodynamic and transport measurements~\cite{ChMa96,CFM95}, which show a return to Fermi-liquid behavior for these concentrations. In contrast, in UCu$_{5-x}$Pd$_x$ $\delta K/(a^{*}\chi_{\rm av})$ at low temperatures is similar for $x = 1.0$ and 1.5~\cite{BMAF96}. The slope~$\delta E/C$ is roughly independent of $x$ in UCu$_{5-x}$Pt$_x$, and a factor of two larger than the slope in UCu$_{5-x}$Pd$_x$.

Why is $\delta E/C$ larger in UCu$_{5-x}$Pt$_x$, $x = 1.5$ and 2.5, if NFL effects are smaller in these alloys? The zero-temperature susceptibility~$\chi(0)$ is a factor of two smaller in UCu$_4$Pt than in UCu$_4$Pd, so that the characteristic energy is a factor of two larger in UCu$_4$Pt. The values of $\delta E/E_{\rm av}$ are comparable (Fig.~2), so that we obtain a larger $\delta E$ and a larger slope $\delta E/C$ for UCu$_4$Pt than for UCu$_4$Pd. For $x({\rm Pt}) = 1.5$ and 2.5 $\chi(0)$ is smaller still, so that even with the larger slope $\delta E/E_{\rm av}$ remains relatively small.

In UCu$_4$Pd we find $\delta E \approx 230$~K from the slope in Fig.~2. This together with Eq.~(\ref{eq:lowtemp}) gives $E_{\rm av} \approx 110$~K, in good agreement with the paramagnetic Curie-Weiss temperature = 133~K from $\chi(T)$~\cite{ChMa96}. In UCu$_4$Pt a similar analysis yields $\delta E \approx 510$~K and $E_{\rm av} \approx 260$~K, in agreement with the paramagnetic Curie-Weiss temperature = 200-300~K from $\chi(T)$ for $x({\rm Pt}) = 1$--2.5. The increase of $E_{\rm av}$ with increasing $x$ is consistent with the observed decrease of the bulk susceptibility at low temperatures~\cite{ChMa96,CFM95}.

In disorder-driven NFL theories the important parameter is $\delta E/E_{\rm av}$. Our results show that the increase of $E_{\rm av}$ in UCu$_{5-x}$Pt$_x$, both in comparison with UCu$_{5-x}$Pd$_x$ and as $x({\rm Pt})$ is increased, produces a decrease in $\delta E/E_{\rm av}$, and hence of NFL behavior, for UCu$_{5-x}$Pt$_x$. The data are consistent with ascribing the NFL behavior of both alloy systems to disorder effects, rather than to the proximity of a quantum critical point in one alloy system but not in the other.

\section*{Acknowledgment}
We are grateful to A. Amato, C. Baines, and D. Herlach for their help during the experiments. One of us (D.E.M.) wishes to thank K. Ishida, Y. Maeno, and the Quantum Materials Group at Kyoto University, and also K. Nagamine and the MSL group at KEK, for their hospitality during visits when part of this work was carried out. These experiments were performed at the Swiss Muon Source, Paul Scherrer Institute, Villigen, Switzerland. The work was supported by US NSF Grants DMR-0102293 (Riverside), DMR-0335173 (San Diego), and DMR-0203524 (Los Angeles), by the U.S. DOE, Contract FG02-04ER46105 (San Diego), and by the Netherlands NWO and FOM (Leiden). Work at Los Alamos was carried out under the auspices of the US DOE.


\begin{thebibliography}{10}
\bibitem{nofit}
We have not attempted a detailed fit of either of these models to the data, in
  view of the difficulties both models have in explaining the low-frequency
  thermal fluctuation spectrum responsible for longitudinal-field muon spin
  relaxation \protect\cite{MHBI04,MBHN01}.
\end{thebibliography}


\end{document}